\documentclass[aps,10pt,twocolumn,prl,floats,floatfix,superscriptaddress]{revtex4-1}
\usepackage{verbatim}
\usepackage{listings}
\usepackage{graphicx}
\usepackage[usenames,dvipsnames,svgnames,table]{xcolor}
\usepackage{amsmath}
\usepackage{amssymb}
\usepackage{bbold}
\usepackage[dvips]{epsfig}
\usepackage[T1]{fontenc}
\usepackage{shadow}
\usepackage{varioref}
\usepackage{epsfig}
\usepackage{array}
\usepackage{rotating}

\newcommand{\be}{\begin{equation}}
\newcommand{\ee}{\end{equation}}
\newcommand{\bea}{\begin{eqnarray}}
\newcommand{\eea}{\end{eqnarray}}
\newcommand{\beqn}{\begin{eqnarray}}
\newcommand{\eeqn}{\end{eqnarray}}
\newcommand{\nn}{\nonumber}
\newcommand{\dG}{\delta G}
\newcommand{\dM}{\delta M}

\newcommand{\td}{\mathrm{d}}

\newcommand{\gmn}{g_{\mu\nu}}
\newcommand{\fmn}{f_{\mu\nu}}

\newcommand{\dd}{\mathrm{d}}
\newcommand{\mpl}{m_\mathrm{Pl}}

\newcommand{\MeV}{\mathrm{MeV}}
\newcommand{\GeV}{\mathrm{GeV}}

\begin{document}
\title{Gravitational origin of Dark Matter}

\author{Eugeny~Babichev} 
\affiliation{Laboratoire de Physique Th\'eorique, CNRS, Univ.~Paris-Sud, Universit\'e Paris-Saclay, 91405 Orsay, France}
\affiliation{UPMC-CNRS, UMR7095, Institut d'Astrophysique de Paris, GReCO, 98bis boulevard Arago, F-75014 Paris, France}

\author{Luca~Marzola}
\affiliation{National Institute of Chemical Physics and Biophysics, R\"avala 10, 10143 Tallinn, Estonia.}
\affiliation{Laboratory of Theoretical Physics, Institute of Physics, University of Tartu; Ravila 14c, 50411 Tartu, Estonia.}

\author{Martti~Raidal}
\affiliation{National Institute of Chemical Physics and Biophysics, R\"avala 10, 10143 Tallinn, Estonia.}
\affiliation{Laboratory of Theoretical Physics, Institute of Physics, University of Tartu; Ravila 14c, 50411 Tartu, Estonia.}

\author{Angnis~Schmidt-May}
\affiliation{Institut f\"ur Theoretische Physik, Eidgen\"ossische Technische Hochschule Z\"urich,
Wolfgang-Pauli-Strasse 27, 8093 Z\"urich, Switzerland}

\author{Federico~Urban}
\affiliation{National Institute of Chemical Physics and Biophysics, R\"avala 10, 10143 Tallinn, Estonia.}

\author{Hardi~Veerm\"ae}
\affiliation{National Institute of Chemical Physics and Biophysics, R\"avala 10, 10143 Tallinn, Estonia.}

\author{Mikael von Strauss}
\affiliation{UPMC-CNRS, UMR7095, Institut d'Astrophysique de Paris, GReCO, 98bis boulevard Arago, F-75014 Paris, France}

\noaffiliation

\begin{abstract} 
Observational evidence for the existence of Dark Matter is limited to its gravitational effects. The extensive program for dedicated searches has yielded null results so far, challenging the most popular models. Here we propose that this is the case because the very existence of cold Dark Matter is a manifestation of gravity itself. The consistent bimetric theory of gravity, the only known ghost-free extension of General Relativity involving a massless and a massive spin-2 field, automatically contains a perfect Dark Matter candidate.  We demonstrate that the massive spin-2 particle can be heavy, stable on cosmological scales, and that it interacts with matter only through a gravitational type of coupling. Remarkably, these features persist in the same region of parameter space where bimetric theory satisfies the current gravity tests. We show that the observed Dark Matter abundance can be generated via freeze-in and suggest possible particle physics and gravitational signatures of our bimetric Dark Matter model.

\end{abstract}

\maketitle

\section{The problem of Dark Matter}\label{sec:mot}

Approximately 85\% of the matter content of the Universe is in the form of Dark Matter (DM), the origin and properties of which still remain unknown.  The existence of DM in our Universe is inferred from its gravitational effects in a number of complementary ways: galactic dynamics (rotation curves and velocity dispersions), gravitational lensing, positions and shapes of the Cosmic Microwave Background peaks, observation of the Baryon Acoustic Oscillations, matter power spectra and simulations of structure formation~\cite{Agashe:2014kda}.

Within the current paradigm, DM is modeled as a cold relic density of an unknown particle produced in the early Universe.  DM models rely on several different production mechanisms, but they usually introduce a new, very weak coupling to baryonic matter. This hypothetical interaction motivates the many current and future dedicated searches aimed at the discovery of DM particles in collider, direct and indirect detection experiments~\cite{Agashe:2014kda, Ackermann:2015zua, Ahnen:2016qkx, Ackermann:2015lka, Khachatryan:2014rra}.

In spite of such extensive effort, DM has thus far remained very elusive, and the experimental null results severely constrain the parameter spaces of viable DM models. Taken at face value, this outcome may in fact point towards the need for a paradigm shift: DM is part of gravity itself and its coupling to Standard Model (SM) particles is suppressed by the Planck mass. In this Letter we demonstrate that such a DM particle is automatically built into the only known consistent extension of General Relativity (GR) to an additional interacting massive spin-2 field.

\section{Ghost-free Bimetric Theory}\label{sec:big}

A recent breakthrough in the physics of gravitation was the construction of ghost-free bimetric theory (see~\cite{Schmidt-May:2015vnx} for a review). This theory contains, in addition to the usual massless graviton, a second propagating spin-2 particle with non-zero mass. Its action describes two dynamical tensor fields $\gmn$ and $\fmn$~\cite{Hassan:2011zd}:
\begin{align}\label{action1}
	S = m_g^{2}\int \td^4 x \Big[&
			\sqrt{|g|}\,  R(g) 
		+ 	\alpha^2 \sqrt{|f|}\,  R(f)\nn\\
		& 	-2m^2\sqrt{|g|}\, V(g^{-1}f)\Big]
		 +S_\mathrm{matter},
\end{align}
where $m_g$ and $\alpha\, m_g $ are the mass scales setting the interaction strengths of the two tensors, while $m$ sets the mass scale for the massive spin-2 field.  The consistency of the theory dictates the form of the potential $V(g^{-1}f)$~\cite{deRham:2010kj,Hassan:2011hr},
\begin{align}
	V\left(\sqrt{g^{-1}f}\right) := \sum^{4}_{n=0} \beta_{n} e_{n}\left(\sqrt{g^{-1}f}\right)\,,
\end{align}
where $\beta_n$ are five free parameters two of which, $\beta_{0}$ and $\beta_{4}$, act as vacuum energy terms for $\gmn$ and $\fmn$ respectively, and $e_{n}(S)$ are the elementary symmetric polynomials of the square-root matrix $S=\sqrt{g^{-1}f}$. They can be defined via the unit weight totally anti-symmetric product,
\be
e_n(S)=S^{\mu_1}_{~[\mu_1}\cdots S^{\mu_n}_{~\mu_n]}\,.
\ee
The absence of ghosts requires that SM matter couples only to one of the metrics in $S_\mathrm{matter}$. Without loss of generality, we will choose the physical metric to be $\gmn$. This then determines the geodesics which SM matter follows and, as we will see, it is in general a mixture of the massless and massive spin-2 modes.

The propagating degrees of freedom of the theory can be read off the action expanded up to quadratic order in the fluctuations $\delta\gmn = \gmn-\bar{g}_{\mu\nu}$ and $\delta\fmn= \fmn-\bar{f}_{\mu\nu}$ around equal backgrounds $\bar{f}_{\mu\nu} = \bar{g}_{\mu\nu}$. These backgrounds correspond to maximally symmetric solutions of the bimetric equations of motion with cosmological constant~\cite{Hassan:2012wr},
\beqn\label{lambda}
	\Lambda
	&=&	m^2\left(\beta_{0} + 3 \beta_1 + 3 \beta_2 + \beta_3\right)\,.
\eeqn
After diagonalization, the quadratic action has the form
\beqn\label{actquad}
&~&
S_{(2)}=\tfrac{1}{2}
\int\dd^4x\Big[\delta G_{\mu\nu} \mathcal{E}^{\mu\nu\rho\sigma} \delta G_{\rho\sigma} + \delta M_{\mu\nu} \mathcal{E}^{\mu\nu\rho\sigma} \delta M_{\rho\sigma} \nn\\
&~&\hspace{70pt} - \tfrac{m_\mathrm{FP}^2}{2}(\delta M^{\mu\nu} \delta M_{\mu\nu}-\delta M^2)\nn\\ 
&~&\hspace{70pt} -\tfrac{1}{\mpl}\Big(\delta G^{\mu\nu}-\alpha\,\delta M^{\mu\nu}\Big)T_{\mu\nu} \Big]\,,
\eeqn
where the kinetic operator $\mathcal{E}_{\mu\nu}^{~~\rho\sigma} \delta G_{\rho\sigma}$ is the linearized Einstein tensor including cosmological constant terms.
The canonically normalized massless and massive eigenstates are, respectively,
\begin{subequations}\label{repmasseig}
\beqn
\dG_{\mu\nu}&=&\tfrac{\mpl}{\sqrt{1+\alpha^2}}\left(\delta\gmn+\alpha^2\delta\fmn\right)\,,\label{repmasseig1}\\
\dM_{\mu\nu}&=&\tfrac{\alpha\,\mpl}{\sqrt{1+\alpha^2}}\left(\delta\fmn-\delta\gmn\right)\,.
\eeqn
\end{subequations}
The quadratic theory then contains a massless graviton $\dG_{\mu\nu}$, which mediates standard gravitational interactions with Planck mass $\mpl \equiv m_g \sqrt{1+\alpha^2}$ and an additional massive spin-2 field $\dM_{\mu\nu}$ whose Fierz-Pauli mass $m_{\rm FP}$ is given by
\be\label{FPmass}
 	m_{\rm FP}^{2} 
	=	m^2\left( 1+ \alpha^{-2}\right)(\beta_{1} + 2\beta_{2} + \beta_{3}) \,.
\ee
Notice that $\alpha$ simply quantifies the mixing between the original metrics $g_{\mu\nu}$ and $f_{\mu\nu}$ and that an overall scale in the $\beta_n$ parameters can be absorbed into the mass scale $m$. Thus, in the following we impose $\beta_{1} + 2\beta_{2} + \beta_{3} = 1$ without loss of generality. 

Following an idea suggested in~\cite{Schmidt-May:2015vnx}, we show here that $\dM_{\mu\nu}$ can behave as a cold DM particle with mass $m_{\rm FP}$. On the other hand, as the particle arises from the gravitational sector, it contributes to gravitational interactions with an effective Planck mass $\mpl/\alpha$. However, for large mass $m_{\rm FP}$, these interactions are exponentially suppressed by the Yukawa shape of the resulting potential and their effect is practically negligible on astrophysical scales.

\section{The GR limits}

In general, bimetric theory introduces modifications to known classical  solutions of General Relativity (GR) at all energy scales, due to the presence of extra propagating degrees of freedom. 
Such modifications are tightly constrained, in particular by Solar System tests of gravity~\cite{Will:2014kxa}. These are usually evaded by invoking the Vainshtein mechanism, which restores General Relativity by means of non-linear self-interactions~\cite{Vainshtein:1972sx}, provided the mass $m_{\rm FP}$ is tuned to tiny values.
In principle, however, there exist two independent parameter limits which restore GR for static solutions in the linear regime (see details in~\cite{Babichev:2013pfa}), namely $m\to\infty$ and $\alpha\to 0$. 
In these parameter regions, bimetric theory automatically passes Solar system tests of GR without invoking the Vainshtein mechanism or any other sort of screening. Moreover, in the case $\alpha \to 0$, all solutions for the physical metric $\gmn$ (not only the static ones) come arbitrarily close to those of Einstein's equations~\cite{Baccetti:2012bk,Hassan:2014vja,Akrami:2015qga}. 
It is also known that instabilities of black holes~\cite{Babichev:2013una} and in cosmological perturbation theory~\cite{Akrami:2015qga} are avoided in the limit of small $\alpha$.
The interesting and nontrivial result that we obtain below is that the massive spin-2 degrees of freedom remain coupled to gravity in both the $m\to\infty$ and $\alpha\to 0$ limits. The massive particle continues to gravitate with the same strength as ordinary baryonic matter and can therefore constitute a suitable DM candidate.

The fact that GR is restored for $\alpha\to 0$ is already suggested by the quadratic action~(\ref{actquad}): in this limit the massive fluctuation $\dM_{\mu\nu}$ decouples from matter and the massless field $\dG_{\mu\nu}$ coincides with the physical metric $\delta\gmn$.  
Notice also that, in principle, a large value for the DM mass $m_{\rm FP}$ in Eq.~(\ref{FPmass}) can be achieved either by suppressing $\alpha$ or by increasing $m$. Because bimetric theory approaches GR even when $m_{\rm FP}$ is held fixed in the $\alpha\rightarrow 0$ limit, we regard $m\to\infty$ as the true limit of infinitely heavy spin-2 field. In the following, $m\to\infty$  will therefore always imply $m_{\rm FP}\to\infty$, while the physical mass $m_{\rm FP}$ is held constant when taking $\alpha\rightarrow 0$.

In order to ensure that our model passes all tests of GR, in this Letter we concentrate on the parameter region where $\alpha\ll1$. The features of the complementary regime characterised by $\alpha\sim 1$ and large values of the mass scale $m$ are briefly discussed below, and will be analysed in detail in a follow-up work~\cite{longversion}.

\subsection{Large spin-2 mass}

Since its formulation, bimetric theory has often been studied in context of the Dark Energy (DE) problem. Hence, the mass scale $m$ of the spin-2 particle is typically assumed to be on the order of the Hubble scale $H_0$. Whereas this assumption avoids fine-tuning the present scale of cosmic acceleration, a value of $m\sim H_0$ is neither a theoretical nor an observational requirement and larger mass values can be considered when the DE problem is addressed in a different way. 
In this case, the $\beta_n$ parameters in (\ref{lambda}) need to be fine-tuned to produce a small~$\Lambda$.

For $m\gg H_0$ the Compton wavelength of the heavy spin-2 is very small, hence the associated nonlinear effects are confined to scales that are inaccessible by current laboratory or astronomical tests of GR.  In fact, bimetric theory introduces modifications to known classical GR solutions in the weak-field linear regime, i.e.\ at large scales, which are suppressed by at least a factor of $\exp(-m_{\rm FP}r)$~\cite{Babichev:2013pfa}.  
This implies that Solar System tests will be automatically satisfied for large values of $m$, corresponding to large $m_{\rm FP}$.  Notice that, in contrast, linear massive gravity with one propagating graviton in the same regime leads to physical predictions different from those of linearised GR~\cite{vanDam:1970vg, Zakharov:1970cc}.

On top of that, the instabilities which generically arise in the cosmological perturbation theory of bimetric theory appear at a much higher energy scale in the large mass limit~\cite{DeFelice:2014nja}. This relegates the associated non-perturbative effects to earlier unobservable cosmological epochs which, as mentioned above, can also be achieved for small values of $\alpha$~\cite{Akrami:2015qga}.

Thus, to summarise, an additional spin-2 field with a large mass is cosmologically viable and yields well-behaved background solutions which satisfy all the Solar System tests of gravity to the current precision.

\section{Validity of perturbative expansion}

One may worry that, for small values of $\alpha$, the theory enters a non-perturbative regime, where the massive mode is strongly coupled. We demonstrate here that this is not the case and that the theory remains weakly coupled within the energy regimes of interest.

The inverse relations between mass and interaction eigenstates in (\ref{repmasseig}) read,
\begin{subequations}\label{invrel}
\beqn
\delta g_{\mu\nu}&=&\tfrac{1}{\mpl}\Big(\delta G_{\mu\nu}-\alpha\,\delta M_{\mu\nu}\Big)\,,\\
\delta f_{\mu\nu}&=&\tfrac{1}{\mpl}\Big(\delta G_{\mu\nu}+\alpha^{-1}\,\delta M_{\mu\nu}\Big)\,.
\eeqn
\end{subequations}
A general vertex of the schematic form $\delta g^k \delta f^n$ in the perturbative expansion of the action around equal backgrounds therefore gives,
\beqn
\delta g^k \delta f^n=\sum_{s=0}^k\sum_{r=0}^n\tfrac{\alpha^{s-r}}{\mpl^{k+n}}{k \choose s}{n\choose r} \delta G^{k+n-r-s}\delta M^{r+s}.
\eeqn
Given that $r\leq n$, every enhancing factor of $\alpha^{-1}$ necessarily appears with at least one suppressing factor of $\mpl^{-1}$. 
Hence, for energies and field values $E\ll\alpha\mpl$, we have a valid double expansion in $\mpl^{-1}$ and $\alpha$. In other words, terms with a higher inverse power $\mpl$ are always suppressed, no matter what their dependence on $\alpha$ is. For terms with the same inverse power of $\mpl$, we can expand in $\alpha$. We stress that, in particular, strong coupling does not arise in the energy regime $E\ll\alpha\mpl$.

These results imply that the cubic vertices deliver the dominant effects of interactions among the massive and massless spin-2 field since they give the correction to the quadratic action (\ref{actquad}) to leading order in $\mpl^{-1}$. We discuss their physical interpretation in the next section whereas their explicit form is provided in Ref.~\cite{longversion}. 

Notice that there is an ambiguity in the definition of the mass eigenstates, connected to the freedom of performing field redefinitions. This issue is discussed in Ref.~\cite{Hassan:2012wr} and more details are provided in Ref.~\cite{longversion}. In particular, when defining the eigenstates $\delta G,\delta M$ in~\eqref{repmasseig}, we could in principle add terms nonlinear in $\delta g, \delta f$. In this case the quadratic $\delta g, \delta f$ interactions would contain cubic interactions for $\delta G,\delta M$. In the following we remove this ambiguity by retaining the linear relations given in Eq.~\eqref{repmasseig} and obtain the cubic interaction vertices whose coefficients are listed in Table~\ref{table1}.
\begin{table}
\begin{center}
\begin{tabular}{c|c|c|c}
$\delta G^3$~ & ~$\delta G^2 \delta M$~ & $\delta G \delta M^2$ & $\delta M^3$ \\
\hline 
&&&\\
 ~$1, \Lambda$~~ & $0$~ & ~$1, \Lambda, m_{\rm FP}^2$ ~& ~$\alpha, \alpha \Lambda,\alpha m_{\rm FP}^2$, $\tfrac{1}{\alpha}, \tfrac{\Lambda}{\alpha}, \tfrac{m_{\rm FP}^2}{\alpha}$~\\
 &&&
\end{tabular}
\end{center}
\caption{Coefficients of cubic interaction vertices (numerical factors neglected) in units of $\mpl^{-1}$. Vertices with a dimensionless coefficient are associated to two derivatives.}
\label{table1}
\end{table}

\section{Phenomenology}
\label{sec:dm}

\subsection{DM interactions}

Let us discuss the effects of each kind of cubic vertex separately, identifying $\dM$ as our DM candidate.

The $\dG^3$ terms are simply the usual gravitational self-interactions arising from the Einstein-Hilbert term of GR, whereas the self-interactions of the massive spin-2 field are given by the $\dM^3$ terms. From Table~\ref{table1} it is clear that some of these vertices are enhanced in the limits of small $\alpha$ or large $m_{\rm FP}$, as compared to the $\dG^3$ terms. Notice however that they are still suppressed by inverse powers of $m_{\rm Pl}$ when compared to SM self-interactions.

The $\delta G^2 \delta M$ terms describe the decay of the massive spin-2 field into two massless gravitons. While this decay would na\"ively be allowed, we find that no such term is present and, therefore, DM does not decay into massless gravitons. The decay into SM fields is still allowed, although it is suppressed by the Planck mass of the matter coupling in Eq.~(\ref{actquad}) as we will discuss in more detail below. We would like to emphasize that in our setup the weakness of the interaction between DM and SM fields descends naturally from the very large value of the physical Planck mass $\mpl$: this is exactly what one expects if DM is a manifestation of gravity itself.

The $\delta G \delta M^2$ terms reveal that the DM field responds to the massless spin-2 field in the same way as standard baryonic matter. Consequently, the massive spin-2 field gravitates exactly as the postulated DM component of $\Lambda$CDM. Remarkably, this coupling is independent of $\alpha$ and the feature persists in the GR limit of small $\alpha$.

The last two points can also be understood by comparing the Noether and gravitational stress-energy tensors in our setup. The Noether stress-energy is computed in the usual way from the quadratic theory, Eq.~(\ref{actquad}) which does not explicitly contain $\alpha$, but only $\Lambda$ and $m_{\rm FP}$. Furthermore, since the quadratic theory is diagonalized, there are no mixing terms $\delta G \delta M$ in the Noether stress-energy. The gravitational stress-energy on the other hand is obtained by varying the cubic interaction terms with respect to $\delta G$. It is known that these two definitions of stress-energy tensor coincide in flat space, i.e.~for $\Lambda=0$, after imposing the equations of motion; see~e.g.~\cite{Leclerc:2005na}. 
This is consistent with the vertices displayed in the second and third column of Table~\ref{table1}, which verify the independence of $\alpha$ as well as the absence of $\delta G^2 \delta M$ terms. Of course, the agreement of the two stress-energy tensors can also be verified explicitly from the exact cubic vertices provided in~\cite{longversion}.
It is also important to highlight that the equivalence between Noether and gravitational stress-energy implies that in the non-relativistic limit the massive spin-2 field acts as a dust source for the massless field.

\subsection{DM decay}

The universal interaction of spin-2 DM with the SM matter allows for its decay into species lighter than $m_{\rm FP}/2$, thereby providing possible signatures for indirect detection experiments. We estimate the associated  decay width into a relativistic species $X$ as~\cite{Han:1998sg}
\begin{align}
	\Gamma(\delta M \to XX) = \frac{C_{X}}{80\pi}\tfrac{\alpha^{2} m_{\rm FP}^{3}}{m_{\rm Pl}^{2}} 
\end{align}
where $C_{X}=\tfrac{1}{6},\tfrac{1}{2} ,1$ for scalars, fermions and gauge bosons, respectively. The constraints on the individual decay widths are heavily dependent on the mass and the decay channels of the DM candidate~\cite{Ibarra:2013cra}. The weakest upper bound on the mass $m_{\rm FP}$ comes from imposing that the decay width into SM particles is less than the inverse age of the Universe; this translates to the limit,
\begin{align}
	\alpha^{2/3} m_{\rm FP} \lesssim 0.1 \,\GeV.  
\end{align}
The most conservative constraint comes instead from Fermi-LAT bounds on the photon flux~\cite{Ackermann:2015zua, Ahnen:2016qkx, Ackermann:2015lka}, which imply $\Gamma(\delta M \to \gamma\gamma) \lesssim 10^{-27}s^{-1}$.  In this case we obtain $\alpha^{2/3}m_{\rm FP}  \lesssim 0.1 \,\MeV$.

As for the possible production mechanisms, the weakness of the Planck-suppressed coupling hints at the possibility of out-of-equilibrium thermal production. In particular, our spin-2 DM can be produced via $s$-channel processes initiated by SM particles and mediated by the massless graviton. Assuming an averaged cross section times velocity of the typical order of $\langle \sigma v\rangle \approx \mpl^{-4} T^2$ at the temperature $T$, matching the observed DM abundance $\Omega_{\rm DM}$ via freeze-in means~\cite{Tang:2016vch},
\begin{equation}
	m_{\rm FP} \approx \frac{\Omega_{\rm DM} \mpl^3}{\Omega_{\rm b} T_*^3} m_{\rm p} \eta_{\rm b}\,,
\end{equation}
where $m_{\rm p}$ is the proton mass, $\Omega_{\rm b}$ the abundance of baryons, $\eta_{\rm b}$ the baryon asymmetry and $T_*$ the maximal reheating temperature. If we require that $T_*$ does not exceed the inflation scale currently indicated by experiments, $10^{14}$ GeV, this implies TeV $\lesssim m_{\rm FP}\lesssim 10^{11}$\,GeV.

\subsection{Gravitational DM signatures}

The most immediate prediction of our proposal is that DM will not be detected in current and future direct and collider searches, simply because its coupling to SM matter is by far too weak. Nonetheless, there are unique signatures which can attest our claim.

Self-interactions of our spin-2 DM are enhanced by inverse powers of $\alpha$. In cluster collisions, baryonic and dark matter would then experience different drag forces possibly resulting in configurations like the one observed in the Abell 520 clusters. Currently, observation of Galaxy cluster mergers yield an upper bound on DM self-interactions of the order of $\sigma_{\rm DM}/m_{\rm DM} \lesssim 1\,{\rm barn}/\GeV$ \cite{Harvey:2015hha}, which however is poorly constraining. Finally, we remark that large DM self-interactions could result in differences between the baryonic and DM power spectra on small scales. 

Another notable property of our DM candidate is that its gravitational interactions may differ from that of SM matter in curved spacetime. While in flat space the Noether and gravitational stress-energy tensors always coincide, the nonlinear mixing of the massive spin-2 field with the graviton could induce different behaviours in the presence of background curvature.
This feature already manifests itself via a rather nontrivial presence of $\Lambda$ in the $\delta G \delta M^2$ terms, c.f.~Table~\ref{table1}. Close to black holes or on cosmological scales, it could then be possible to detect modified gravitational interactions of DM. 

Of course, our framework could be falsified by investigating additional signatures of bimetric theory which are not related to DM phenomenology. 
For instance, one possibility lies in observations of black holes. Indeed, since the standard no-hair theorem does not apply to bimetric theory, it is natural to expect that black holes, in general, possess hairs formed by the absorption of spin-2 DM particles.
Another option is provided by the fact that the interaction term for the metrics $\gmn$ and $\fmn$ introduces corrections to Friedmann's equation which affect both expansion history and cosmological perturbation theory (see \cite{Solomon:2015hja} for a summary). Depending on the values of $\alpha$ and $m$, i.e.~on how close the theory is to GR, these effects can be observable as deviations from $\Lambda$CDM.

\section{Conclusions}\label{sec:con}

We have identified an ideal DM candidate in the only known ghost-free extension of GR which includes a massive spin-2 field: the massive spin-2 particle is stable on cosmological scales and its interactions with SM fields are very weak.
Remarkably, our DM particle possesses standard gravitational interactions in flat space and in parameter regions where bimetric theory passes all observational tests. In other words, bimetric theory resembles GR plus a gravitating DM particle, the origin of which is purely gravitational. 
The weakness of the interactions between DM and SM fields arises naturally from the weakness of gravitational interactions or, equivalently, from the large value of the physical Planck mass.

Observational signatures of our DM candidate range from indirect detection experiments (due to DM decay) to the observation of possible DM self-interactions in cosmic mergers. Assuming a thermal freeze-in DM production mediated by the massless graviton constrains the DM mass to the range TeV $\lesssim m_{\rm FP}\lesssim 10^{11}$\,GeV. 

\vspace{5pt}

{\bf Note added:} In the final stages of this work, Ref.~\cite{Aoki:2016zgp}, which overlaps with this work, appeared on the arXiv.

\vspace{5pt}

\acknowledgments
We thank A.~Hektor for discussions.
This work was supported by the Russian Foundation for Basic Research Grant No. RFBR 15-02-05038 (EB), by the ERC grants IUT23-6, PUTJD110, PUT 1026 and through the ERDF CoE program (LM, MR, FU, HV), by ERC grant no.~615203 under the FP7 and the Swiss National Science Foundation through the NCCR SwissMAP (ASM) and by the ERC grant no.~307934 under the FP7/2007-2013 (MvS).

\end{document}